\documentclass[conference]{IEEEtran}
\ifCLASSINFOpdf
   \usepackage[pdftex]{graphicx}
\else
   \usepackage[dvips]{graphicx}
\fi
\begin{document}
\title{Review on the Advancements of DNA Cryptography}

\author{\IEEEauthorblockN{Beenish Anam}
\IEEEauthorblockA{School of Computing\\
The Bradford University\\
West Yorkshire, UK\\
b.anam@brad.ac.uk}
\and
\IEEEauthorblockN{Kazi Sakib}
\IEEEauthorblockA{School of Computing\\
The Bradford University\\
West Yorkshire, UK\\
k.muheymin-us-sakib@bradford.ac.uk}
\and
\IEEEauthorblockN{Md. Alamgir Hossain}
\IEEEauthorblockA{School of Computing\\
The Bradford University\\
West Yorkshire, UK\\
M.A.Hossain1@Bradford.ac.uk}
\and
\IEEEauthorblockN{Keshav Dahal}
\IEEEauthorblockA{School of Computing\\
The Bradford University\\
West Yorkshire, UK\\
k.p.dahal@Bradford.ac.uk}
}


%Telephone: (800) 555--1212, Fax: (888) 555--1212}


\maketitle

\begin{abstract}
Since security is one of the most important issues, the evolvement of cryptography and cryptographic analysis are considered as the fields of on-going research. The latest development on this field is DNA cryptography. It has emerged after the disclosure of computational ability of  Deoxyribo Nucleic Acid (DNA). DNA cryptography uses DNA as the computational tool along with several molecular techniques to manipulate it. Due to very high storage capacity of DNA, this field is becoming very promising. Currently it is in the development phase and it requires a lot of work and research to reach a mature stage. By reviewing all the potential and cutting edge technology of current research, this paper shows the directions that need to be addressed further in the field of DNA cryptography.
\end{abstract}

\section{Introduction}
\label{intro}
DNA cryptography, a new branch of cryptography utilizes DNA as an informational and computational carrier with the aid of molecular techniques. It is relatively a new field which emerged after the disclosure of computational ability of DNA~\cite{adleman94}. DNA cryptography gains attention due to the vast storage capacity of  DNA, which is the basic computational tool of this field. One gram of DNA is known to store about $10^8$ tera-bytes. This surpasses the storage capacity of any electrical, optical or magnetic storage medium ~\cite{cui06,chen03}.

DNA is being proposed to use for many computational purposes. For example, Barish et. al. demonstrated a tile system that takes input and produces output using DNA~\cite{barish05}. The method is now also used to solve many NP-complete and other problems. Such as Rothemund et. al. showed that DNA can also be used to compute XOR function which is an essential part of  cryptosystems~\cite{rothemund04}.
It is a very potential field of research, as work which has been done in this field suggests that it can put many challenges to the modern cryptosystems~\cite{cui08}. By utilizing DNA cryptography, several methods have been designed to break many modern algorithms like Data Encryption Standard (DES)~\cite{boneh95}, RSA~\cite{beaver94,brun07} and  Number Theory Research Unit (NTRU)
~\cite{pelletier02, zhang08}.

The research of DNA cryptosystem is still in its early stage. Thus, the scope of doing research on this new field is multi-dimensional. Work needs to be done from theory to realization, as both of the dimensions yet to be matured. Recent development showed that some key technologies in DNA research, such as Polymerase Chain Reaction (PCR), DNA synthesis, and DNA digital coding, have only been developed~\cite{tanaka05}.

Traditional cryptographic systems have long legacy and are built on a strong mathematical and theoretical basis. Traditional security systems like RSA, DES or NTRU are also found in real time operations. So, an important perception needs to be developed that the DNA cryptography is not to negate the tradition, but to create a bridge between existing and new technology. The power of DNA computing will strengthen the existing security system by opening up a new possibility of a hybrid cryptographic system. 

This paper gives a simple comparison between traditional and DNA cryptographic methods. It gives an insight to the benefits which can be achieved with the help of DNA cryptography and discusses the techniques which are currently used in this field. It also stresses on the need that both the traditional and DNA cryptographic techniques should be merged in a way that resulting cryptographic systems can enjoy the benefits from both the fields. It also points out some loop holes in this field and discusses that this field needs further research to gain the stage of realization.

The rest of this paper is organized as follows:  Section~\ref{brw} gives background knowledge which is required to understand DNA cryptography and an insight is given to the field of cryptography, DNA and DNA cryptography. In this section, a typical cryptographic scenario is also explained, which is used in later sections. Section~\ref{bio_prob} discusses the techniques which are used in DNA cryptography. What has been done and what still needs to be done will be discussed in Section~\ref{sor}. Finally, conclusion is drawn in Section~\ref{con}.

\section{Background and Related Work}
\label{brw}
Modern cryptography uses cross disciplinary interactions between mathematics, computer science, and engineering. Applications of cryptography include online banking, computer authetication, and e-commerce. The discussion starts with the most basic approach for cryptography, and then improvements are demonstrated.

\subsection{Cryptography}
Cryptography is a very significant and widely used field, because it is the basis of security of all information. It is a very old field, but in modern times due to the increased use of internet, the significance of this field has been multiplied. The whole manual systems most importantly banking, defence and shopping systems are now being converted into web applications. The precious data which is being transferred over the internet is vulnerable to many security attacks~\cite{galbreath02} such as IP spoofing, man in the middle attack, teardrop etc. To secure our systems and applications, we rely on the strength of cryptography.

Cryptanalysis runs parallel to the cryptography.  The aim of the cryptanalysis is to analyse and try to break the security systems proposed by the discipline of cryptography~\cite{menezes96}. So that means, how strong a cryptographic system is, depends on how weak cryptanalysis is possible for that system~\cite{gehani04}. A great deal of work has been done on the cryptography and cryptanalysis. And as a result, various systems have been designed (for example, RSA, ECC, etc.) to achieve high level of security.

\subsection{Cryptographic Scenario}
The typical scenario in cryptography is that Alice (sender) wants to send some messages secretly to Bob (intended receiver). The message which is to be sent is in the ordinary language understood by all, is called a plaintext.  The process of converting plaintext into a form which cannot be understood without having special information is called encryption. This unreadable form is called cipher text and the special knowledge is called encryption key. 

The conversion of cipher text again into plaintext with a special knowledge is called decryption, whereas special knowledge for decryption is called decryption key. Only the receiver has this special knowledge and only receiver can decrypt a cipher text with this knowledge called decryption key. In traditional cryptography encryption and decryption is done by algorithms for which currently there is no available solution~\cite{cui08}.

\subsection{Types of Cryptography}

There are three prominent branches or sub fields of cryptography~\cite{cui08}, named as: 
\begin{enumerate}
	\item Modern Cryptography
	\item Quantum Cryptography
	\item DNA Cryptography.
\end{enumerate}

These three fields depend upon different difficult problems concerning to different disciplines for which there is no known solution until now. The modern cryptography is based upon the difficult mathematical problems such as prime factorization, elliptic curve problem, for which there is no known solution found so far. Quantum cryptography which is also relatively a new field, is based upon the Heisenberg uncertainty principle of Physics, and DNA cryptography depends upon the difficult biological processes concerning to the field of DNA technology~\cite{cui08} such as :
\begin{itemize}
	\item Polymerase Chain Reaction (PCR) for a sequence without knowing the correct two primer pairs~\cite{cui08, tanaka05}
	\item and another is extracting information from the DNA chip without having the knowledge about the sequences present in different spots of DNA chip~\cite{mingxin07}.
\end{itemize}

A detailed explanation of these technqiues is given in Section 3.

\subsection{DNA}
\begin{figure}
	\centering
	\includegraphics[width=3.0in]{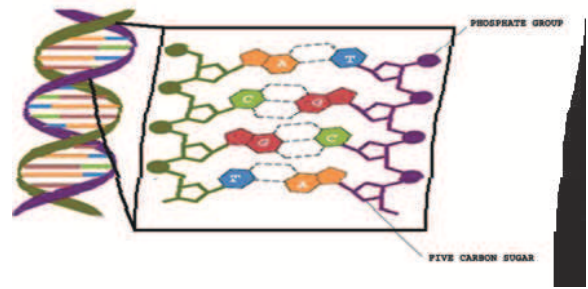}
	\caption{Structure of DNA (A - Adenine, G - Guanine, C - Cytosine and T - Thiamine).}
	\label{dnapix}
\end{figure}

Deoxyribo Nucleic Acid (DNA) is  the hereditary material of almost  entire living organisms ranging from very small viruses to complex human beings~\cite{lister10}. It is an information carrier of all life forms. DNA is a double helical structure with two strands running anti parallel as shown in Figure~\ref{dnapix}. DNA is a long polymer of small units called nucleotides. Each nucleotide consists of three components:
\begin{enumerate}
	\item a Nitrogenous base
	\item a five carbon sugar
	\item a Phosphate group
\end{enumerate}
There are four different nucleotides depending upon the type of nitrogenous base they have got. There are four different bases A, C, T, G called Adenine, Cytosine, Thiamine and Guanine respectively~\cite{zhang04, watson53}.

DNA stores all the huge and complex information about an organism with the combination of only these four letters A, C, T and G. These bases form the structure of DNA strands by forming hydrogen bonds with each other to keep the two strands intact. A forms hydrogen bond with T whereas C and G forms bonds with one another~\cite{guozhen06}. It can be seen in Figure~\ref{dnapix}. 

Until 1994, DNA was believed to carry only the biological information but it was Adleman who revealed the computational ability of DNA when he solved NP complete Hamiltonian path problem of seven vertices~\cite{adleman94}. After that DNA has been used as a computational tool as well~\cite{leier00}. DNA computers deal with the DNA language that consists of four letters A, C, T and G~\cite{lister10, zhang04}.
The computational ability of DNA is now used in cryptography as well. DNA cryptography is a very potential field and if manipulated in appropriate manner can give much harder competition to other fields of cryptography~\cite{cui09}.

\begin{table*}
\caption{Comparison between traditional and DNA cryptographic methods}
\begin{center}
\label{bg_tab1}
\begin{tabular}{|l|c|c|c|c|c|}
\hline
\hline
&&&&&\\
\textbf{}&\textbf{Security}&\textbf{Time Complexity}& \textbf{Storage Medium} & \textbf{Storage Capacity} & \textbf{Stability}\\
&&&&&\\
\hline
&&&&&\\
\textbf{Traditional Cryptography} & One Fold & $\geq$ few seconds & Computer (Silicon) Chips & 1 gram of silicon chip & Dependant on \\
&&&& carries 16 MB~\cite{farnel10} & implementation environments\\
&&&&&\\
\hline
&&&&&\\
\textbf{DNA Cryptography} & Two Fold & $\geq$ few hours & DNA strands & 1 gram of DNA & Dependant on \\
&&&& carries $10^8$ TB~\cite{cui09} & environmental conditions\\
&&&&&\\
\hline
\hline
\end{tabular} 
\end{center}
\end{table*}

\subsection{DNA Cryptography}

Cryptography is the science that addresses all the aspects for secure communication over an insecure channel, namely privacy, confidentiality, key exchange, authentication, and non-repudiation. As mentioned above, DNA provides an excellent mean to secure data, the technique has been named as DNA cryptography. In such techniques, plaintext message data is encoded in DNA strands by the use of an alphabet of oligonucleotide sequences. Natural DNA obtained from biological sources may be recoded using nonstandard bases, to allow for subsequent processing~\cite{adleman94,lister10}. Input and output of the DNA data can be moved to conventional binary storage media by DNA chip arrays. Where binary data may be encoded in DNA strands by use of an alphabet of short oligonucleotide sequences.

In Table~\ref{bg_tab1}, a simple comparison is shown between traditional and DNA cryptographic techniques by considering attributes including security provided by the technique, time taken to process the technique, storage medium which is used to store the data in the technique, the storage capacity of the storage medium used and the stability of the results for the particular technique.

By considering  security provided by both the techniques, it can be seen that DNA cryptographic techniques provide two fold security by involving computational difficulties as well as the biological difficulties. In Traditional Cryptography security can be said to be one fold as it relies on only computational difficulties.  Time taken by the efficient cryptographic algorithms is few seconds, whereas DNA cryptographic techniques that involve PCR and DNA chip technology can take hours to complete the whole process. Traditional cryptography generally runs on computers over the network, so the storage mediums are silicon chips of the computers, whereas DNA cryptography deals with the DNA strands which are manipulated by biological techniques. If we consider DNA as the storage medium, it has got huge storage capacity as compared to the equivalent amount of the silicon chips. This property of data makes DNA cryptography and DNA computing very tempting and beneficial field of research.

Stability of results of cryptographic techniques refers that encryption and decryption always gives the same results. If the stability of the results of encryption and decryption provided is analysed, it can be seen that traditional cryptographic algorithms depends on the implementation conditions~\cite{jabri01}. Implementation conditions include the platform and the language limitations used to encode the algorithm.  Whereas the stability of DNA cryptography is very much dependant on the environmental conditions such as temperature, pH. This instability of DNA cryptography is discussed later. 
%In summary, a basic network information sharing approach was initially suggested for organising sensor network topologies. Due to high energy overhead of network information sharing, different self-organising techniques have been proposed such as clustering~\cite{heinzelman02,taek03,sohrabi04}, hierarchy~\cite{cheng03,ma04} or adaptive topology~\cite{cerpa02,cerpa04}. Although these approaches can reduce network control overhead, they produce uneven energy consumption of nodes by only considering individual node energy usage. Uneven node energy usage resulted in unevenly distributed node lifetime. As a result, the early exhaustion of an important node degrades the network quality of service such as coverage and connectivity.

\section{Difficult Biological Problems Used In DNA Cryptography}
\label{bio_prob}

DNA cryptography utilizes biological methods for encryption and decryption. Among those, Polymerase Chain Reaction (PCR)~\cite{cui08, tanaka05} and DNA chip technology~\cite{mingxin07} are the most prominent cryptographic techniques. However, steganography using DNA, is also found in the literature. All these techniques are described below.

\subsection{Polymerase Chain Reaction (PCR)}
PCR is an amplification and quantification process of DNA.  The purpose of designing PCR is to increase the amount of DNA, as it is very difficult to deal with small amount of DNA strands. The name Polymerase chain reaction comes from the enzyme (biological catalyst) known as polymerase used in the technique and chain represents that this amplification process occurs in many cycles one after another. By performing PCR, short sequences of DNA can be analysed even in samples containing only minute quantities of DNA. PCR can select small strands of DNA and amplifies those. In practice, amplification of DNA involves cloning of segments of interest into vectors for expression. PCR is highly efficient so that untold numbers of copies can be made from small selected DNA. Moreover, PCR uses the same molecules that nature uses for copying DNA. To perform PCR, one should know the sequence of DNA to be amplified to design the right primer for it, where primer is a sequence containing few numbers of nucleotides complimentary to the specific  sequence of DNA which is to be amplified~\cite{cui08}. In short, we can identify the PCR process into two phases:

\begin{itemize}
	\item Two "primers", short single-stranded DNA sequences to correspond to the beginning and ending of the DNA stretch.
	\item Polymerase enzyme that moves along the segment of DNA, reading its code and assembling the copy.
\end{itemize}

\begin{figure}[!t]
\centering
\includegraphics[width=3.5in]{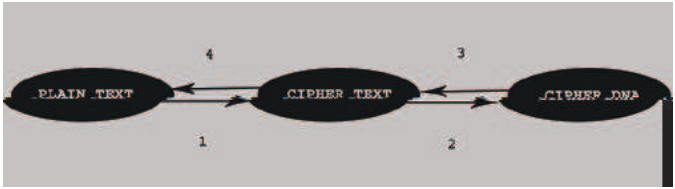}
\caption{Summary of Cryptographic Technique using PCR}
\label{cipher}
\end{figure}

%\subsubsection{Cryptographic Techniques using PCR}
The encryption key, in this case, is compound, consisting of both PCR primers pair and public key.  Similarly decryption key consists of complementary primer pairs and private key. Encryption starts with the exchange of two primers (forward and reverse) between Alice and Bob via a secure channel~\cite{cui08}. For Encryption, pre-processing can be done, that the whole algorithm like RSA can be applied first. This step is numbered as 1 in Figure~\ref{cipher}. Then cipher text can be converted into DNA sequence by coding scheme represented by number 2 in Figure~\ref{cipher}. By performing this, entirely different cipher text can be obtained~\cite{cui08}. In literature, cipher DNA refers the term cipher text which is in the form of DNA sequence, and plain text DNA denotes the plain text which is in the form of DNA. 

The prepared cipher DNA is then flanked by the secret primers and mixed with a number of other unknown DNA.  Alice sends this DNA mixture to Bob~\cite{cui08}. For decryption, Bob can retrieve the cipher DNA by performing PCR using its secret primer, and reverse the whole process which is done for encryption~\cite{cui08}. Anyone without knowing the two primers cannot retrieve target cipher DNA~\cite{tanaka05}. Decryption steps are represented as 3 and 4 in Figure~\ref{cipher}. In the same figure, number 3 denotes that the cipher DNA can be retrieved by using DNA decryption key (secret primers) and is converted into cipher text by using coding scheme. Finally, in the figure, 4 denotes that the cipher text is decrypted by RSA private key. 

The above mentioned PCR technique has numerous different implications, such as Guangzhao Cui et. al. proposed a encryption scheme with the aid of PCR amplification and DNA coding schemes~\cite{cui08}. On the other hand, Tanaka et. al. devised a public-key system using DNA as a one-way function for key distribution using PCR amplification to restore the plaintext DNA from other distractor DNA strands~\cite{tanaka05}. Yamamoto et. al. also contributed in cryptography by producing large-scale DNA memory based on the nested PCR~\cite{yamamoto08}. This technique of DNA cryptography provides two fold security by involving both the molecular techniques and modern algorithms. If one security level is broken some way, the other can keep this technique safe~\cite{cui08}. However, the problem of PCR based techniques lies in the transportation of secret keys between the sender and receiver. Which has been discussed in detail in Section~\ref{sor}.  

%\subsubsection{Security Level}

\subsection{Steganography using DNA}

Steganography is the technique of hiding information. The goal of cryptography is to make data unreadable by a third party; on the other hand, the goal of steganography is to hide the data from a third party. Formula shown at equation~\ref{stg_eq} is a very generic description of the steganographic process:

\begin{eqnarray}
	cover_{medium} + hidden_{data} + key_{stego} = stego_{medium}
\label{stg_eq}
\end{eqnarray}

In this context, the $cover_{medium}$ is the file in which the data, $hidden_{data}$, is hid. The resultant file may also be encrypted using another key called $key_{stego}$. Finally, the $stego_{medium}$ is the file that will be transported. Usual, medium of covering file may include the audio or image file. However, due to its massive storage capability, DNA is getting popular to be a steganographic covering medium. 

The primitive idea of  DNA steganography can be described as follows. For encryption, one or more input DNA strands are taken to be tagged as the plaintext message. One or more randomly constructed secret key strands are appended with the input DNA. Resulting "tagged plaintext" DNA strands are hidden by mixing them within many other additional "distracter" DNA strands which might also be constructed by random assembly. For decryption, given knowledge of the "secret key" strands, resolution of DNA strands can be decrypted by a number of possible known recombinant DNA separation methods. Such as plaintext message strands may be separated out by hybridization with the complements of the "secret key" strands might be placed in solid support on magnetic beads or on a prepared surface. These separation steps may combined with amplification steps.

One such method is found in~\cite{risca01}. Viviana I. Risca proposed a steganographic technique using DNA and standard biological protocols. The proposed method encodes the information in the sequence of  DNA strand, flanked by two secret primer target regions. The technique uses monoalphabetic encryption key to assign random but unique 3-base DNA codons to 40 alpha-numeric and punctuation characters. The key was then used to encode plaintext messages into the base sequence of an artificially synthesized oligonucleotide.

The resultant DNA strand is then hidden into a very large amount of similarly sized garbage DNA strands. To extract the message, one must know the primers that will bind the target regions on the message-containing DNA strand in order to selectively amplify the required molecule. 

Since, this primitive technique is simple and easily breakable, an improvement is proposed by Asish Gehani et. al~\cite{gehani04}. The idea is to distinguishing the probability distribution of plaintext source from that of distracter DNA stands~\cite{gehani04}.

As it is discussed that DNA based steganography  converts plaintext into plain text DNA  but do not encrypt it. It only hides the plaintext DNA into bulk of other DNA and person who knows the primer for it can easily locate the plaintext DNA and amplify it. This technique can be very useful as it saves the cost of encryption, but it is vulnerable to statistical analysis. So the PCR based cryptography may be considered to be much safer~\cite{cui08, cui09}.

\subsection{DNA Chip Technology}

A brief introduction to DNA chip technology is given to clear the concept of this technique used in cryptography. 
Over less than a decade, DNA chips and microarrays have changed the way in which scientists carry out their investigations. DNA chips enable researchers to manipulate the vast amounts of data from genome-sequencing~\cite{gwynne01}. DNA chip technology is very important for the manipulation of biological data. It is commonly used to find expression of many genes in parallel~\cite{zhang08}. These chips like silicon chips can be used to handle and store the data in the form of DNA sequences. 

DNA chips consist of large number of spots embedded on a solid surface, most commonly used is a glass slide. Each spot consists of different kind and number of probes, where probes are small nucleotide sequences which are able to bind to the complimentary nucleotides.  Nucleotides which bind to these probes are fluorescently labelled, whenever any DNA sequence binds to these probes, it is observed under a laser dye and data is calculated electronically depending upon the ratio of the binding of probe with the DNA in each spot~\cite{tsukahara04, brown99}. Manufacturers are now developing smaller biochips with better information handling capabilities that will contribute more effectively to numerous fields of research including cryptography.

Technique, considering the typical cryptographic scenario, has following steps:

\begin{itemize}
	\item Encryption key is a collection of particular probes where decryption key is a collection of corresponding probes having complimentary sequence. The decryption key is then sent to the Bob in a secure manner.
	\item Plaintext is converted into a binary format. This binary format is then embedded into DNA chip as a cipher text (cipher DNA). Without knowing the decryption key one cannot read the plaintext from the DNA chip.
	\item Bob uses the decryption key and hybridizes the cipher DNA. With the help of a computer software he can retrieve plain text~\cite{mingxin07}.
\end{itemize}

One such cryptosystem is XOR One-time-pad~\cite{gehani04}. To construct one-time pad using DNA chip technology, an array of immobilized DNA strands are used, where multiple copies of a single sequence are grouped together in a microscopic pixel. The strands are optically addressable. Distinct DNA sequences at each optically addressable site of the array can be synthesised using various methods. One of the known such technologies is combinatorial synthesis conducted in parallel at thousands of locations. For preparation of oligonucleotides of length L, the $4^L$ sequences are synthesized in 4n chemical reactions.

To encrypt, each plaintext message has been appended with a unique prefix index of length $L_0$. A complement of the plaintext message tag is created on  one-time-pad DNA sequence by appending unique prefix index tag of the same length $L_0$. Using annealing and ligation each of the corresponding pair of plaintext message and a one-time-pad sequence is concatenated. After that the message is encrypted by bit-wise XOR mechanism. XOR operation is shown in Equation~\ref{chip_eq1}, where $C$ is the cipher data or strands, $M$ is the plaintext message and $S$ is a sequence of independently distributed random bits.

\begin{eqnarray}
	C_i = M_i \otimes S_i for = 1,…,n.
\label{chip_eq1}
\end{eqnarray}

However, the fragments of the plaintext are converted to cipher strands and plaintext strands are dropped. For decryption commutative property of bit-wise XOR operation is used, as shown in Equation~\ref{chip_eq2}-\ref{chip_eq3}.

\begin{eqnarray}
C_i \otimes S_i &=& (M_i \otimes S_i) \otimes S_i \\
\label{chip_eq2}
&=& M_i \otimes (S_i \otimes S_i) \\
&=& Mi.
\label{chip_eq3}
\end{eqnarray}

MingXin et. al. used DNA chip technology in a similar manner as described above to design a symmetric key encryption scheme, they referred this encryption scheme as DNA symmetric-key cryptosystem (DNASC)~\cite{mingxin07}. Recently Lai et. al. proposed an asymmetric encryption scheme using DNA chip technology and they also designed signatures using this DNA chip technology so that Alice and Bob can verify each other~\cite{xuejea10}.

The encryption technique using DNA chip technology is not restricted to encrypt only textual data, there are also encryption schemes which are designed to encrypt and decrypt images. One such cryptosystem is designed by  Gehani  et. al. where they produced a substitution one-time-pad system and used it during the process of encryption and decryption of 2D images. It is noteworthy that they also performed PCR to amplify particular DNA strands~\cite{gehani04}. Similar technique is also noticed in Shyam et. al.'s work, where they proposed a cryptosystem and created plaintext and cipher text pairs using DNA chip technology but on images~\cite{shyam08}. 

Although DNA chip provides a wide range of parallel data processing capability, the lack of interoperability of DNA chip and other storage mediums supressed its potentiality at this moment.

\section{Scope of Research}
\label{sor}

There are many advantages which seem to be associated to the field of DNA cryptography. The huge storage capacity of DNA makes it a very tempting field for research. Moreover the cryptographic techniques which are designed by involving this field are believed to give very high security level~\cite{cui08}.

The research which has been done so far on DNA cryptography shows that several DNA-based methods can be devised in order to break cryptosystems which are currently being used. Many cryptosystems used today are based on RSA public key encryption. RSA public key encryption is based on the intractability of prime factorization as there are no known efficient algorithms to find the prime factors of sufficiently large numbers. As shown in Equation~\ref{prime_eq} -
\begin{eqnarray}
	n=p*q
	\label{prime_eq}
\end{eqnarray}

where p and q are prime numbers, for a given "n" it is infeasible to find p and q when n is a very large number. If there is any technique that can find how to factor given "n", the whole RSA scheme will be broken. There are techniques which have been devised to break RSA scheme in DNA cryptography. These techniques used self-assembly of DNA tiles to fully break RSA scheme~\cite{brun08, brun07}. If these techniques are able to break RSA, RSA will no more remain practical.  

Elliptic curve cryptography (ECC) has been applied in key exchange and also in the digital signatures. The security of these cryptosystems is based on the difficulty of solving the elliptic curve discrete logarithm problem~\cite{miller85}. DNA-based methods have been developed to break the cryptosystems based on elliptic curves. These methods are accomplished by means of basic biological operations and have developed a parallel multiplier; a parallel divider and a parallel adder for adding points on elliptic curves~\cite{li08}.

The encryption schemes (Ciphers) that uses keys only once are said to be one-time pads (OTP). In theory OTP cipher is absolutely secure. But practically, key distribution and key generation are critical issues to be resolved for the use of OTP ciphers. Key space should be large enough so that keys can only be used once. DNA having huge storage capacity, can be manipulated to generate key space to be used for OTP cipher. 

There are also some areas that need to be improved. For example,  Time and computational complexity are two of the most important parameters for any kind of cryptographic systems, DNA cryptography dealing with the manipulation of DNA sequences  takes a lot of time to deal and work out with DNA sequences as compared to time taken by many very efficient algorithms of traditional cryptography such as, DES, RSA etc.~\cite{beaver94,brun07}. 

There is also a risk associated to DNA cryptography that if the cipher DNA gets impured by unwanted DNAs, this process may collapse. This may be handled by taking precautions and maintaining the laboratory environment pure. Still research must be carried on to address the reliability issues.

To summarise the whole discussion, following points are significant. 

\begin{enumerate}
	\item In cryptographic technique involving PCR technology, it can be examined that there are two encryption keys and two decryption keys. One pair of encryption and decryption key is (n , e) pair which was used for RSA public key cryptosystem. The other pair was the forward and reverse primers used for tagging the cipher DNA. These primers were necessary to be securely shared among Alice and Bob as DNA can only be recovered if these primer pairs were known. This means that in addition to the public and private keys, primer pairs are also to be secretly transported. In public key encryption there is only one secret which is shared that is the Bob's private key. However in this technique primer pairs are also to be secretly transported~\cite{cui08}. The transport of primers is not as simple as for the transport of keys of the traditional cryptographic techniques. 
	 
Primer pairs are the sequences of nucleotides; there is a probability that these molecules can be affected by environmental conditions. To take account of this, environmental conditions should be maintained throughout the process of transport. 
	 
Another possibility is that the sequence of primer pair in the form of A,C,T,G can be sent as digital form to the Bob, same as the manner in which keys of traditional cryptographic techniques are transported. Then Bob can synthesize his own primer pairs in the laboratory considering the sequence which is sent to him by Alice. During all this process environmental conditions should remain stable.
		\item DNA steganography system can be the complement of the traditional DNA cryptography, however, potential limitations of DNA steganography methods is its vulnerability. Certain DNA steganography systems can be broken, with some assumptions on information theoretic entropy of plaintext messages. May be the method is brute force, but cannot be ruled out because of the advancement of computing technologies. To make the situation better, plaintext messages can be compressed before encrypting, however, in this case, plaintext messages need to be pre-processed. So the open question is whether DNA steganography systems with natural DNA plaintext input can or cannot be made to be unbreakable. 
	
The process of DNA steganography involves the tagging of plaintext DNA with the primers and is hid in other garbage or junk DNA of same length as the plaintext DNA. This is a very simple method to provide confidentiality. Same issue can arise here if the environmental conditions change the binding property of DNA with the primer. There is also a possibility that other garbage DNA can bind to the primer, making it difficult to recover the original plaintext DNA~\cite{cui08,cui09}.
	
	\item DNA chips contains arrays of Pico moles of  DNA sequences, known as probes. Since an array can contain huge number of probes, a DNA chip can accomplish number of computation in parallel. However, DNA chip data is difficult to exchange between traditional storage medium, due to the lack of standardization in fabrication, protocols, and analysis methods~\cite{fuscoe06}. This problem is defined as interoperability problem in bioinformatics. Moreover by examining the DNA chip technology used in the DNA cryptography, it is seen that encryption and decryption processes are performed on DNA chip. DNA being a biological molecule; whose properties are dependent on the environmental conditions. For instance, DNA's property to bind to other nucleotides can be changed by changing environmental conditions. So encryption and decryption process does not remain stable. Due to this instability, encryption and decryption may show  different results under different environmental conditions.
\end{enumerate}

%Key transportation is an inevitable process in all symmetric or asymmteric cryptosystems. All the cryptosystems rely on the fact that keys cannot be forged during the process of encryption, decryption and also during their transport. It can be suggested that this field should be analysed for the transportation of keys securely. Key transportation is a very critical step in the field of cryptography. Techniques can be used to transport private keys which then can be used for traditional cryptographic algorithms to encrypt or decrypt the data. In this case, small amount of data would be handled, so does increase the efficiency. Security of these keys is more important than the time taken to transport the keys. Sometimes, time can be compromised but security cannot be.

In short it can be said that DNA cryptography cannot totally be replaced traditional cryptography which is currently being used. This field requires a lot of research and work to have a position in which it can be implemented and used for practical purposes. There is a need that people from traditional cryptography and DNA technology should exchange knowledge among each other and cryptosystems should be devised in such a way that they can enjoy benefit from both the fields.

\section{Conclusion}
\label{con}

DNA cryptography is a relatively new cryptographic field of research evolved with the DNA computing. In this particular field, DNA is used as message carrier and the bio-technology such as PCR, is used as the implementation mechanism. The extravagant storage capability and parallel computability of DNA molecules are exploited for encryption, authentication, and authorization. In this paper, the existing DNA cryptographic techniques namely Polymerase Chain Reaction (PCR), Steganography using DNA and DNA chip technology have been discussed. With the summarization of the progress of DNA cryptographic research, the advantages, future trends and several problems have also been identified. All the three kinds of cryptography have their own advantages and disadvantages and can be treated as the complement of each other in future security applications. However, the difficulties that are identified in DNA cryptography are the absence of theoretical basis and practical methodologies which can readily be implemented in the field of security.

The potentiality of DNA on computing can open up further biological molecule based computation methods. Once the DNA cryptography field is developed and analysed, attempts can be made to convert the cipher DNA into cipher proteins or RNA, these can give us another level of security. It will be possible only by intensive research and practical work on DNA computing.

\section*{Acknowledgment}
The Erasmus Mundus project eLINK - east-west Link for Innovation, Networking and Knowledge exchange programme (EM ECW SGA2008-4949) for funding this study.

%\bibliography{mybiobibfile}
\bibliographystyle{IEEEtran}

\end{document}